 \documentclass[preprint]{aastex}

\shortauthors{Moon, Pirger, \& Eikenberry}
\shorttitle{High-Speed Data Acquisition System}

\begin{document}


\title{A Next Generation High-speed Data Acquisition System 
for Multi-channel Infrared and Optical Photometry}








\author{Dae-Sik Moon, Bruce E. Pirger\altaffilmark{1}, and Stephen S. Eieknberry}
\affil{Department of Astronomy, Cornell Universiy, Ithaca, NY 14853;
moon,pirger,eiken@astrosun.tn.cornell.edu}
\altaffiltext{1}{http://www.acqsys.com, Acquisition Systems, Langmuir Lab Box 1002, 95 Brown Road, Ithaca, NY 14850}


\begin{abstract}
We report the design, operation, and performance of a next generation
high-speed data acquisition system for multi-channel infrared and optical
photometry based on the modern technologies of
Field Programmable Gate Arrays,
the Peripheral Component Interconnect bus,
and the Global Positioning System.
This system allows either direct recording of
photon arrival times
or binned photon counting with time resolution up to 1-$\mu$s precision
in Universal Time,
as well as real-time data monitoring and analysis.
The system also allows simultaneous recording of multi-channel
observations with very flexible, reconfigurable observational modes.
We present successful 20-$\mu$s resolution simultaneous observations of
the Crab Nebula Pulsar in the infrared (H-band) and
optical (V-band) wavebands obtained with this system
and 100-$\mu$s resolution V-band observations of the dwarf nova
IY Uma with the 5-m Hale telescope at the Palomar Observatory.
\end{abstract}

\keywords{infrared: general --- instrumentation: miscellaneous --- nova, cataclysmic variables --- pulsars: individual (Crab Nebula Pulsar) --- stars: individual (IY UMa) --- techniques: photometric}


\section{Introduction}

Real-time, high-speed (microsecond time resolution)
data acquisition systems for multi-channel
infrared and optical photometry have wide potential applications
in modern astronomy, not only for scientific observations
but also for development of new observational technology.
These include observations of compact objects
(such as pulsars, X-ray binaries, and cataclysmic variables) 
(e.g., Eikenberry, Fazio, \& Ransom 1996),
observations of planetary and lunar occultations 
(e.g., French et al. 1996; Simon et al. 1999),
fringe detection for infrared and optical interferometry 
(e.g., Mallbet et al. 1999),
and wavefront sensing for adaptive optics systems 
(e.g., Colucci et al. 1994).

Out of the various applications mentioned above, 
some require precise knowledge of exact Universal Time (UT).
For instance, recording of precise UT time is always 
required in pulsar observations for comparison with 
the pulse profiles previously obtained 
and/or obtained from other wavelengths,
and for studying phase coherence of pulsations.
Observations of the planetary and lunar occultations also 
require the UT time information.
In addition, many recent important astronomical discoveries have been
possible through the simultaneous multi-wavelength observations
(e.g., Eikenberry et al. 1998), which also require precise UT time information
for comparison of the data from different observatories.
On the other hand, since astronomers use the same telescope, 
but different detectors, for infrared and optical observations,
it is highly desirable that the data acquisition system is capable of
accepting multi-channel inputs, 
allowing simultaneous high-speed photometry 
in infrared and optical wavebands with the same telescope.

Previous data acquisition systems used for infrared and
optical high-speed photometry are typified by
the ``Lil Wizard Pulsarator," built by Richard Lucinio
at Caltech and Jerome Kristian at the Carnegie Observatory
in the early 1980s.
The Wizard has two sets of two counting registers. One register
accumulates counts from a detector, while the other
counts a preset number of oscillations from a frequency standard.
This system is capable of counting photons with time resolution
between 5-$\mu$s and 64-ms
and transfering data directly to an Exabyte tape device,
along with header information including the number of
frequency oscillations since power-up.
The Wizard, however, does not allow
real-time data monitoring and analysis
nor direct recording of the UT arrival time of each photon,
which makes its applications very limited and inefficient.

The technologies of digital electronics and computer interfaces
have developed so fast recently that it is now possible to build
an inexpensive next generation multi-channel high-speed data acquisition system
which allows real-time data monitoring and analysis as well as
direct recording of UT photon arrival times.
In this paper we describe the
Cornell High-Speed Data Acquisition System (CHISDAS)
for multi-channel infrared and optical photometry,
based on the technologies of 
Field Programmable Gate Arrays (FPGAs), 
the Peripheral Component Interconnect (PCI), and the
Global Positioning System (GPS).
Through the FPGA technology,
CHISDAS is a fully reconfigurable system,
offering very flexible observational modes,
including binned counting with time resolution
up to 1 $\mu$s.
Direct recording of the UT arrival time of each photon
is also possible up to 1-$\mu$s precision
by decoding of Inter Range Instrumentation
Group Time Code Format B (IRIG-B) signal
from a high-precision GPS receiver.
CHISDAS is designed to accept signals from up to 4 data channels
simultaneously, and can be easily reconfigured to accept
signals from more channels.

\section{Hardware and Overall Performance}

A simplified block diagram of CHISDAS hardware in two-channel mode is shown in Fig. 1,
where the infrared and optical photometers based on 
the Solid-State Photomultiplier (SSPM) and Photomultiplier Tube (PMT)
are used as typical high-speed photometers for the given wavebands, respectively.
Analog signals from the photometers
are first brought into a Stanford Research SR400 Photon Counter via BNC
cables, digitized, and output again via BNC to a
Signal Conditioning Box which converts the input 
signals from the SR400 to TTL digital logic signals.
The Signal Conditioning Box also accepts 
1- and 10-MHz frequency standards,
amplitude modulated IRIG-B signals, 
and CMOS 1-pulse-per-second (1PPS) absolute time reference pulses
from the GPS receiver.
The frequency standards and the IRIG-B signals are converted 
into TTL pulses in the Signal Conditioning Box.
All output signals from the Signal Conditioning Box are then
connected to a data acquisition card inside a Pentium PC 
through a Very High Density Cable Interconnect (VHDCI) cable.

\begin{figure}
\plotone{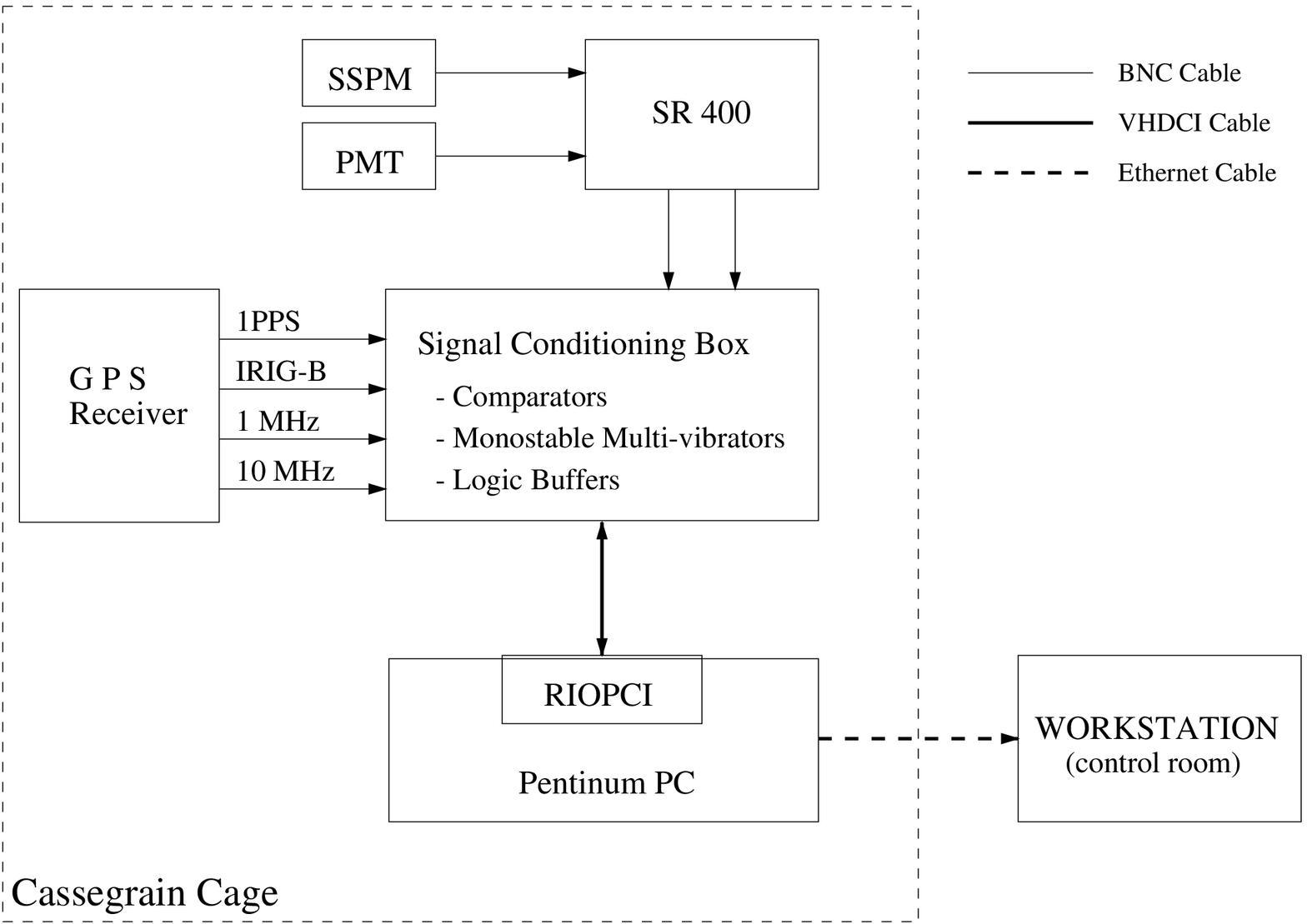}
\caption{A simplified block diagram of CHISDAS in two-channel mode. \label{fig1}}
\end{figure}

\placefigure{fig1}

Fig. 2 shows a simplified block diagram of the Signal Conditioning Box.
The negative going pulses (from 0 to $-$0.7 V) from the SR400
are converted to TTL pulses with a simple comparator circuit
(with $-$350 mV reference level) and a monostable
multivibrator. (The monostable multivibrators are used in series with
comparators to avoid possible retriggering of the comparators.)
The 1- and 10-MHz frequency standards and the IRIG-B signals are converted
to TTL pulses in the same way but with reference levels of
+0 and +1.5 V, respectively.
High-speed logic buffers are finally used to drive
all the input signals through the VHDCI cable
to the RIOPCI card.

\begin{figure}
\plotone{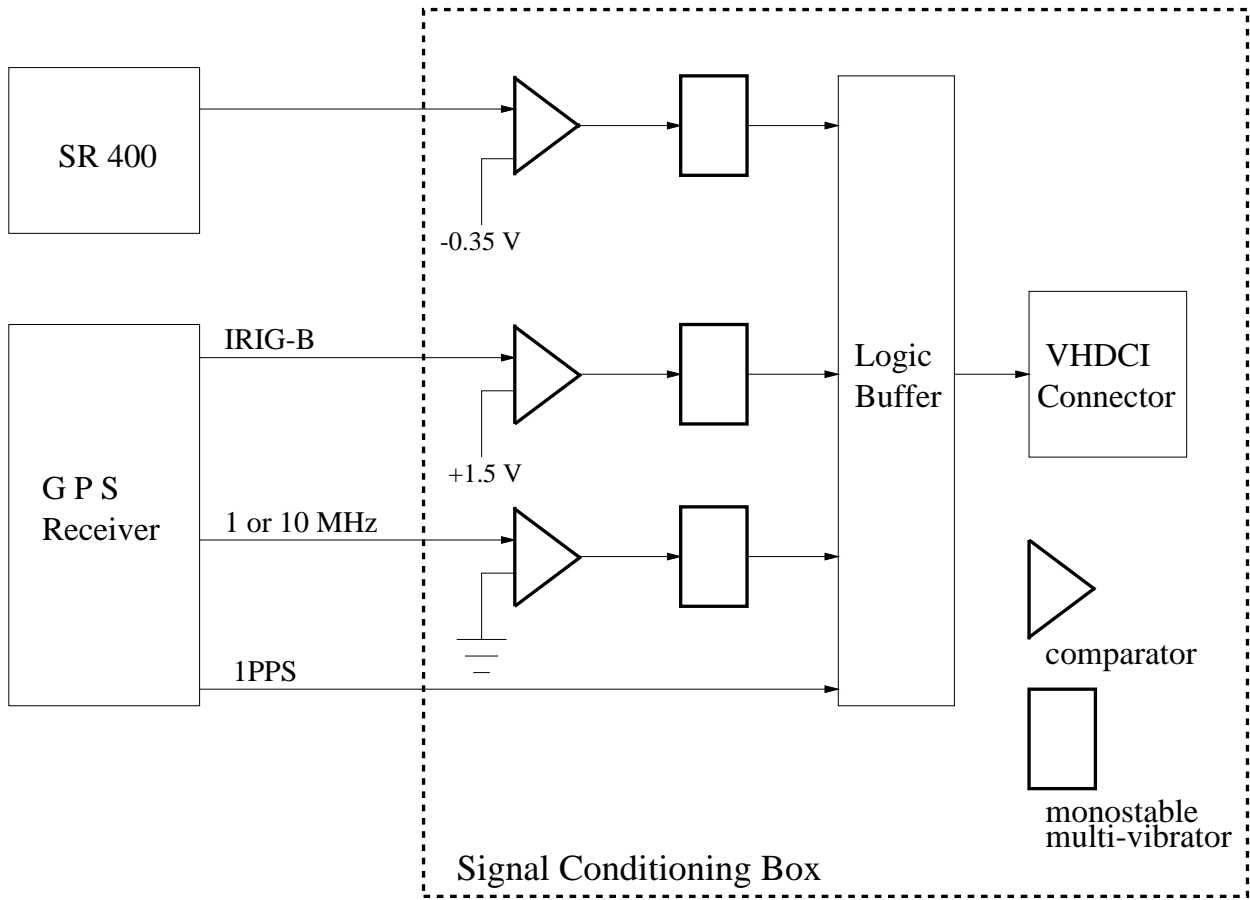}
\caption{A block diagram of Signal Conditioning Box.
Simple circuits of comparators and monostable multivibrators
are used to conver input analog signals from SR400 and GPS receiver
to digital logic signals.
Logic buffers are used to drive signals through VHDCI cable.
\label{fig2}}
\end{figure}

\placefigure{fig2}

For the data acquisition card,
we use the RIOPCI card provided by 
Acquisition Systems\footnote{http://www.acqsys.com}.
This card runs on the Windows NT operating system,  
utilizing an Altera FPGA, high-speed SRAM,
an AMCC S5933 ASIC interface to the PCI bus, 
an integrated daughter card interface,
and a flexible clocking system.
The most important part of the RIOPCI card for CHISDAS application
is the FPGA which provides digital logic resources,
such as logic gates and registers,
that can be interconnected by digital designs
to create a digital logic for CHISDAS application.
The FPGAs are software configured and can be easily reconfigured 
to different modes with different digital designs.
All of the RIOPCI card resources are available to the end user 
for custom application development.

The heart of CHISDAS is the digital logic implemented in
the FPGA on the RIOPCI card.
The digital logic is based on the hierarchical structure of state machines
which have several interconnected parts 
including an interface to the PCI interface, 
a part for the IRIG-B signal decoding, 
a multi-channel data counter,
an interface to the SRAM, 
and an interface to the daughter card. 
The FPGA is first programmed to accept an instruction from 
the observer through the PCI interface. 
The instruction is a ``double-word" (DWORD; a 32-bit digital signal)
containing information for UT start time of observations, time resolution, 
data size for bus-master transfer, and number of input data channels. 
(The communication between the observer and the FPGA on the RIOPCI
card are carried out by C programs interfacing the observer and 
the registers of the PCI interface. 
We shall call these C programs ``CONTROLLER."  See \S 3 for detailed explanations.)
In order to start binning of input photon pulses 
exactly from the given start time of observations,
the FPGA is programmed to compare 
the input IRIG-B signals from the GPS receiver 
with the given start time of observations 
from the instruction DWORD.
The photon pulses from the detectors are then binned
to the given time resolution with reference to the 
1- or 10-MHz GPS frequency standards.

The IRIG-B signal from the GPS receiver is a 
binary-coded-decimal amplitude-modulated 1-kHz sine wave.
The logic-high amplitude is +2.5$\pm$0.5 V while the
logic-low is +0.75$\pm$0.1 V.
It has information for the exact UT time of the incoming 1PPS pulses.
A simple comparator circuit
with a monostable multivibrator is used to convert the 
binary-coded-decimal wave to TTL pulses.
The TTL pulses are then decoded in FPGA programs
according to the format of the IRIG-B signal\footnote{http://jcs.mil/RCC/files/200.pdf}
to determine the exact UT time of the incoming 1PPS.

The resulting counted value of each bin
is then buffered to the local SRAM in the RIOPCI card.
The housekeeping data are also buffered to the local SRAM
once per record (one record corresponds to 8192 bins where 
one bin corresponds to the given time resolution).
The housekeeping data are five DWORDs including 
(1) a security DWORD marking the beginning of the housekeeping data,
(2) the instruction DWORD given by the observer,
(3) a 40-bit ``master-counter" value clocked by frequency standards from the GPS receiver, 
(4) a 24-bit ``record-counter" value, and
(5) a 32-bit ``bin-counter" value.
The 40-bit master-counter value corresponds to the total integrated frequency standard 
clocks from the start of observations. 
The 24-bit record-counter value corresponds to the record number from the start
of observations, while the 32-bit bin-counter value corresponds to the 
bin number within the given record.
The final data are, therefore, a series of 8192 binned values of each record 
following the housekeeping data of 5 DWORDs.
The buffered data are then
directly transferred to the memory of the PC 
by means of bus-master transfer (i.e., direct memory access).
Meanwhile new data are continuously written to the local SRAM in a different SRAM region.
The size of each bus-master transfer is given by the observer through the instruction.

Once in PC memory, the CONTROLLER first checks the housekeeping data 
in order to confirm the continuity of the transferred data.
For any incoming errors in the housekeeping data, the CONTROLLER generates
error messages and stops operation of the RIOPCI card.
After examining the housekeeping data, the CONTROLLER 
writes the data saved in the PC memory onto a file on local hard disk.
It finally transfers the file 
over an Ethernet connection to the hard disk of the workstation 
in the telescope control room.
The transferred data are then analyzed by real-time IDL programs on the workstation
(we shall call these programs ``ANALYZER", see \S 3)
and recorded on tapes for storage.

\section{Software}

The CHISDAS software consists of two parts: the CONTROLLER 
on the PC and the ANALYZER on the workstation.

Three main functions of the CONTROLLER are
(1) controlling the PCI interface on the RIOPCI card 
to configure the FPGA,
to give instructions to the FPGA through the FIFO, 
and to save the transferred data, via bus-master transfer from the FPGA, in the PC memory,
(2) monitoring the incoming housekeeping data to examine the operation of the CHISDAS,
and (3) transferring the data in PC memory 
to the hard disk on the workstation in the telescope control room through 
an Ethernet connection.

ANALYZER includes programs for: 
(1) plotting photon counts per second for the real-time monitoring of the incoming data, 
(2) making time series from the raw data and rebinning the time series, and
(3) performing Fast Fourier Transformation (FFT) and power spectrum analysis.
It also includes several advanced programs especially useful for pulsar observations,
such as routines for 
(1) barycentering the time series using the JPL 200 ephemeris based on 
    TEMPO\footnote{http://pulsar.princeton.edu/tempo},
(2) Fourier interpolation (Eikenberry 1997),
(3) signal-folding in order to find frequency and its derivative (Eikenberry 1997), and
(4) generating pulse profiles.

\section{Observational Examples}

CHISDAS has been successfully used with infrared and optical high-speed photometers 
based on SSPM and PMT on the 5-m Hale telescope of the Palomar Observatory 
for various sources incluing pulsars, X-ray binaries, and cataclysmic variables.
Two examples of the observations with CHISDAS are given here.
One is 20-$\mu$s simultaneous infrared and optical waveband observations of the
Crab Nebula Pulsar, which is specifically based on CHISDAS capability
of multi-channel observations and direct recording of UT time.
(Note that the Unconventional Stella Aspect [USA] X-ray satellite [Ray et al. 1999],
which has onboard GPS receiver, was used together for simultaneous 
infrared, optical, and X-ray observations.)
The other is 100-$\mu$s V-band observations of the eclipsing dwarf nova
IY UMa (Uemura et al. 2000), for which the $\sim$10$^2$-s time scale variability
of the source (such as eclipse and humps) was well traced in real time 
due to the real-time monitoring capability of CHISDAS.

Fig. 3 shows the $\sim$4-h infrared (H-band) and optical (V-band)
pulse profiles of the Crab Nebula Pulsar obtained 
by the two-channel mode photometry
with 20-$\mu$s resolution on Nov. 28, 1999.

\begin{figure}
\plotone{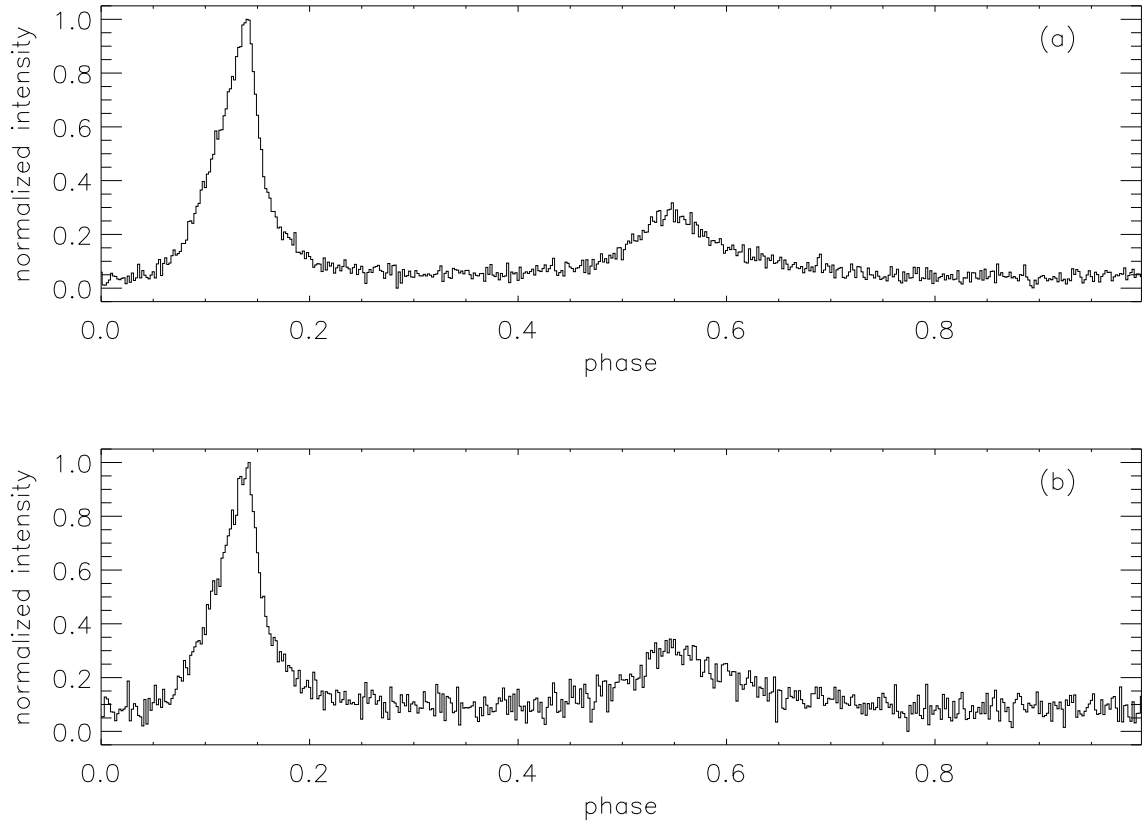}
\caption{(a) Crab Nebula Pulsar V-band pulse profile obtained with 20-$\mu$s resolution.
(b) Same as (a), but for H-band.
The profiles are obtained simultaneously with $\sim$4-h integration.
\label{fig3}}
\end{figure}

\placefigure{fig3}

The observed data were analyzed with ANALYZER.
First, the photon arrival times from the raw data were transformed to
the solar system barycenter using the JPL DE200 ephemeris.
The barycentered time series was then rebinned to 2.56-ms resolution.
The maximum power frequency was found to be $f_{\rm max}$ $\simeq$ 29.847194 Hz
through the FFT of the binned time series. 
The routines for the fourier interpolation and signal folding were used
to find the signal-folded frequency of 
$f_{\rm sf}$ $\simeq$ 29.8472(1)\footnote{the value in the parenthesis is the 1-$\sigma$
uncertainty in the last quoted digit}.
During the signal folding, the frequency derivative was fixed to be the value,
$\dot f \simeq -3.7461751 \times 10^{-10} \; \rm s^{-2}$,
of the monthly Jodrell Bank radio ephemeris
\footnote{http://www.jb.man.ac.uk/$\sim$pulsar/Resources/resources.html}.
The signal folded frequency is in good agreement with the
Jodrell Bank radio freqeuncy, $f_{\rm r}$ $\simeq$ 29.8472117964(2).
Finally, the pulse profiles in Fig. 3 were obtained using the Jodrell Bank ephemeris.
This is the first-ever, as far as we are aware, simultaneous photometry 
in infrared and optical wavebands with high time resolution.
A full analysis of the data will be presented elsewhere with the 
data of the USA X-ray observations.

Fig. 4 shows the V-band lightcurve of the dwarf nova IY UMa of which orbital period
is $\sim$1.77 h (Uemura et al. 2000).
The 100-$\mu$s resolution raw data were binned to 4.096-s resolution, 
and we can clearly identify several features including a gradual rise before the eclipse,
a deep eclipse, and two post-eclipse superhumps.
These are unique observations providing detailed information of the source,
since the previous studies rely on the observations 
with small telescopes (e.g, 40-cm aperture)
and with poor time resolutions (e.g 30 s), 
which make it impossible to study the source in detail (see, Uemura et al. 2000).
A full analysis of the data will be presented elsewhere.

\begin{figure}
\plotone{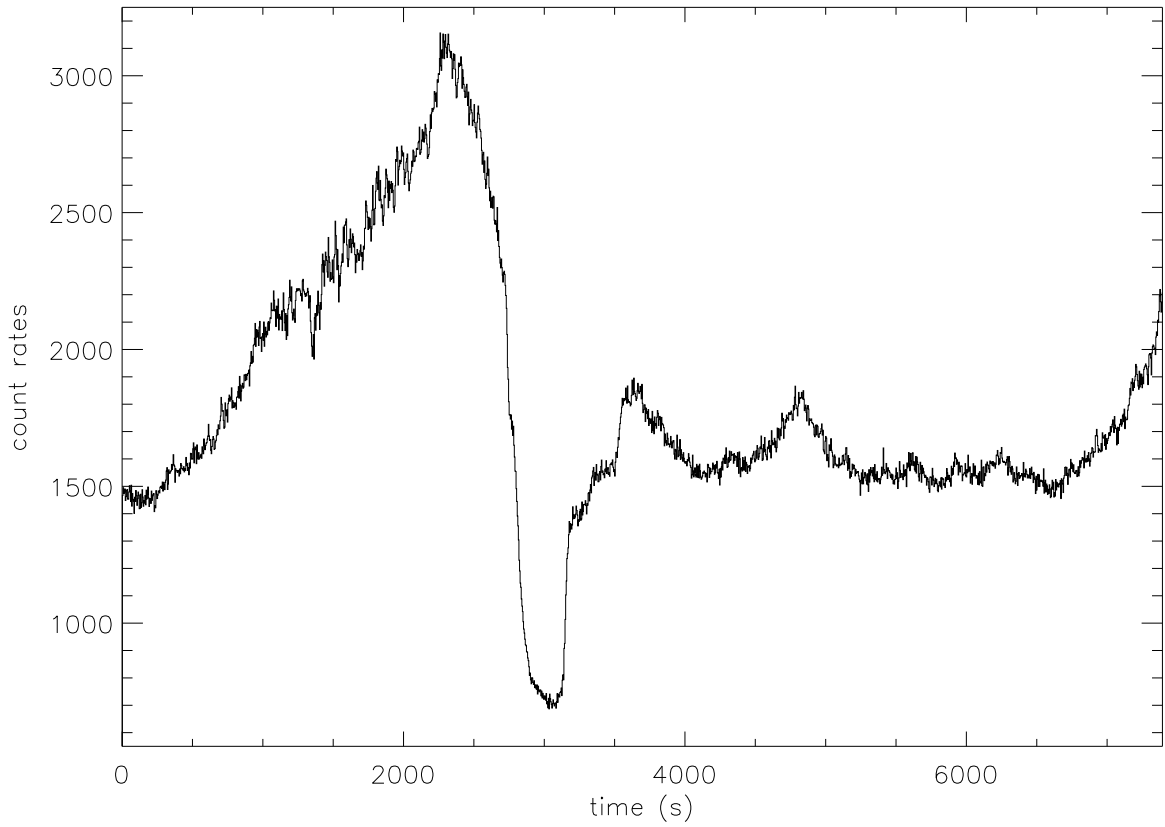}
\caption{V-band lightcurve of the dwarf nova IY UMa obtained by binning the
100-$\mu$s raw data to 4.096-s resolution.
\label{fig4}}
\end{figure}

\placefigure{fig4}

\section{Conclusions}
We have described a next generation
high-speed data acquisition system for multi-channel
infrared and optical photometry.
CHISDAS opens a new window for high-speed infrared and
optical photometry in many ways. Among them are
(1) real-time data monitoring and analysis,
(2) direct recording of UT photon arrival times,
(3) simultaneous observations in multi-channels, and
(4) full reconfigurability of the system.
CHISDAS has already proved to be very powerful and successful
in real observations, and will be used widely 
for high-speed infrared and optical photometry.

\acknowledgments

We would like to thank the referee and editor for comments
and suggestions which simplified the appearance of this paper.
We thank Justin Schoenwald for his help on the AMCC S5933 PCI
device driver from Acquisition Systems. We also thank Chuck Henderson
and Craig Blacken for their help on building the Signal Conditioning Box.
We finally thank the staff members of the Palomar Observatory
for their help on our observations.
DSM is supported by NSF grant AST-9986898.
SEE is supported in part by an NSF Faculty Early Careeer Development
(CAREER) award (NSF-9983830).

\clearpage

\clearpage

\clearpage

\clearpage



\begin{thebibliography}{}
\bibitem[Colucci et al.\ 1994]{cetal94}
    Colucci, D., Lloyd-Hart, M., Wittman, D., Angel, R., Ghez, A., \& McLeod, B. 1994, PASP, 106, 1104
\bibitem[Eikenberry, Fazio, \& Ransom \ 1996]{efr96}
    Eikenberry, S. S., Fazio, G. G., \& Ransom, S. M. 1996, PASP, 108, 993
\bibitem[Eikenberry\ 1997]{eik97}
    Eikenberry, S. S. 1997, Ph.D. thesis, Harvard Univ.
\bibitem[Eikenberry et al.\ 1998]{eet98}
    Eikenberry, S. S., Matthews, K., Morgan, E. H., Remillard, R. A., \& Nelson, R. W. 1998, ApJ, 494, L61
\bibitem[French et al.\ 1996]{fetal96}
    French, R. G., et al. 1996, Icarus, 119, 269
\bibitem[Mallbet et al.\ 1999]{metal99}
    Mallbet, F., Kern, P., Schanen-Duport, I., Berger, J.-P., Rousselet-Perraut, K., \& Benech, P. 1999, A\&AS, 138, 135
\bibitem[Ray et al.\ 1999]{ret99}
    Ray, P. S., et al. 1999, preprint (astro-ph/9911236)
\bibitem[Simon et al.\ 1999]{setal99}
    Simon, M., Beck, T. L., Greene, T. P., Howell, R. R., Lumsden, S., \& Prato, L. 1999, AJ, 117, 1594
\bibitem[Uemura et al.\ 2000]{uet00}
    Uemura, M., et al. 2000, PASJ, 52, 9
\end{thebibliography}
\end{document}